\documentclass{emulateapj}
\usepackage{units,times,natbib}
\usepackage{amsmath}

\def\farcmin{\hbox{$.\mkern-4mu^\prime$}}

\def\earth{\hbox{$\oplus$}}

\def\arcsec{\hbox{$^{\prime\prime}$}}

\def\solar{\mbox{$_{\normalsize\odot}$}}

\def\x{$\times$}
\def\ap{$\approx$~}
\def\hst{{\em HST }}

\def\farcm{\hbox{$.\mkern-4mu^\prime$}}
\def\farcs{\hbox{$.\!\!^{\prime\prime}$}}
\def\earth{\hbox{$\oplus$}}

\def\arcsec{\hbox{$^{\prime\prime}$}}
\def\solar{\mbox{$_{\normalsize\odot}$}}

\newcommand{\lsim}{\ \raise
-2.truept\hbox{\rlap{\hbox{$\sim$}}\raise5.truept\hbox{$<$}\ }}
\newcommand{\gsim}{\ \raise
-2.truept\hbox{\rlap{\hbox{$\sim$}}\raise5.truept\hbox{$>$}\ }}

\slugcomment{Accepted for Publication in ApJ}

\shorttitle{The IMF of NGC~602 in the SMC with HST/ACS} 
\shortauthors{M. Schmalzl et al.}

\begin{document}


\title{The Initial Mass Function of the Stellar Association NGC~602 in
the Small Magellanic Cloud with Hubble Space Telescope ACS
Observations\altaffilmark{1}}

\altaffiltext{1}{Research supported by the Deutsche
Forschungsgemeinschaft (German Research Foundation)}


\author{Markus Schmalzl, Dimitrios A. Gouliermis}
\affil{Max-Planck-Institute for Astronomy, K\"onigstuhl 17, 69117
Heidelberg, Germany;\\ schmalzl@mpia.de, dgoulier@mpia.de}

\author{Andrew E. Dolphin}
\affil{Raytheon Corporation, 870 Winter Street Waltham, MA
02451, USA; adolphin@raytheon.com}

\and

\author{Thomas Henning}
\affil{Max-Planck-Institute for Astronomy, K\"onigstuhl 17, 69117
Heidelberg, Germany; henning@mpia.de}


\begin{abstract}

We present our photometric study of the stellar association NGC~602 in
the wing of the Small Magellanic Cloud (SMC). The data were taken in
the filters $F555W$ and $F814W$ using the {\em Advanced Camera for
Surveys} (ACS) on-board the {\em Hubble} Space Telescope (HST). 
Photometry was performed using the ACS module of the stellar photometry
package DOLPHOT.  We detected more than 5,500 stars with a magnitude
range of 14~\lsim~$m_{555}$~\lsim~28~mag. Three prominent stellar
concentrations are identified with star counts in the observed field,
the association NGC~602 itself, and two clusters, one of them not being
currently in any known catalog. The Color-Magnitude Diagrams (CMDs) of
both clusters show features typical for young open clusters, while that
of the association reveals bright main sequence (MS) and faint
pre-main sequence (PMS) stars as the members of the system. We construct
the initial mass spectrum (IMS) of the association by applying an
age-independent method of counting the PMS stars within evolutionary
tracks, while for the bright MS stars we transform their magnitudes to
masses with the use of mass-luminosity relations. The IMS of NGC~602 is
found to be well represented by a single-power law, corresponding to an
Initial Mass Function (IMF) of slope $\Gamma \approx -1.2$ for
1~\lsim~$M$/M{\solar}~\lsim~45. This indicates that the shape of the IMF
of a star forming system in the SMC for stars with masses higher than
1~M{\solar} seems to be quite similar to the field IMF in the solar
neighborhood.

\end{abstract}

\keywords{galaxies: star clusters --- Magellanic Clouds --- open
clusters: individual(\objectname{NGC~602}) --- stars: evolution ---
stars: pre-main-sequence stars: luminosity function, mass function}


\section{Introduction}

The Small Magellanic Cloud (SMC), the second closest undisrupted
neighboring dwarf galaxy to our own (after the Large Magellanic Cloud;
LMC) is an ideal laboratory to investigate star formation at an
environment similar to the early universe. Its metallicity is by a
factor of 5 lower than the solar metallicity (Luck et al. 1998; Ven
1999), and its dust-to-gas ratio is measured to be up to 10 times lower
than in the Milky Way (e.g. Stanimirovic et al. 2000). Therefore, the
SMC may be considered as an excellent local template for studying
primordial star formation, providing insight into the processes at work
in the early universe. The recent star formation process in the SMC is
characterized by a rich sample of {\sc H~ii} regions (Henize 1956;
Davies et al. 1976) linked to young stellar systems, the {\em stellar
associations}, known for their early-type stellar content and loose
structure (e.g. Kontizas et al. 1999). The small distance of the SMC
from us ($\approx$~60~kpc) allows us to resolve from space individual
stars down to $M_{V}\approx$~10~mag, and consequently the study of its
low-mass young stellar content and the corresponding Initial Mass
Function (IMF) becomes unusually accessible.

Stellar associations contain the richest sample of young bright stars in
a galaxy. Our knowledge on the young massive stars of the Magellanic
Clouds (MCs) has been collected from ground-based studies of young
stellar associations (see e.g. Massey 2006 and references therein).
However, our knowledge of the low-mass stellar membership and the
corresponding IMF in star-forming regions of the MCs is quite
incomplete. A first attempt to define the low-mass population of a
stellar association in the MCs was made by Gouliermis et al. (2005).  In
their photometric study of the LMC association LH~52 with data from the
{\em Wide-Field Planetary Camera 2} (WFPC2), these authors test the
hypothesis that sub-solar-mass stars {\em can be} detected in the MCs.
They found that all faint main sequence stars, which are observed in the
area of the association, belong to the general field of the LMC and not
to the system. The field-subtracted mass function of LH~52, which
accounts for its IMF was found by these authors with a slope $\Gamma
\sim -1.1$ for main sequence stars down to $M \simeq 1$~M{\solar}. This
slope is comparable, but somewhat more shallow than a typical Salpeter
(1955; $\Gamma \sim -1.35$) IMF. An IMF well reproduced by a power law
with a slope consistent with Salpeter's was also found by Sirianni et
al. (2002) in the central regions (within 30\arcsec) of the young SMC
star cluster NGC~330.

\begin{deluxetable}{ccccc}
\tablewidth{0pt}
\tablecaption{Log of the observations
\label{t:obs}}
\tablecomments{Datasets refer to HST archive catalog. Exposure 
times (T$_{\rm expo}$) per filter are given in seconds. Units of right
ascension are hours, minutes, and seconds, and units of declination are
degrees, arcminutes, and arcseconds.}
\tablehead{\colhead{Visit} & \colhead{RA} & \colhead{DEC} & 
\multicolumn{2}{c}{T$_{\rm expo}$}\\[1ex]
\colhead{Dataset} & \multicolumn{2}{c}{(J2000.0)}& \colhead{F555W} & 
\colhead{F814W}}
\startdata
J92F05 & 01~29~28.21 & $-$73~33~16.7& 430 & 453 \\[.5ex]
       & 01~29~28.52 & $-$73~33~16.4& 430 & 453 \\[.5ex]
       & 01~29~27.89 & $-$73~33~16.9& 430 & 453 \\[.5ex]
       & 01~29~27.57 & $-$73~33~17.1& 430 & 453 \\[.5ex]
       & 01~29~27.57 & $-$73~33~17.1& 3 & 2 \\[.5ex]
       & 01~29~27.25 & $-$73~33~17.4& 430 & 453 \\[2ex]
J92FA6 & 01~29~27.57 & $-$73~33~17.1& 3 & 2
\enddata
\end{deluxetable}

\begin{deluxetable}{lcccccc}
\tablewidth{0pt}
\tablecaption{Sample from the photometric catalog of stars found in this
study in the region of NGC~602/N~90 with HST/ACS imaging
\label{t:catalog}}
\tablehead{
\colhead{}& 
\colhead{R.A.}& 
\colhead{DECL.}&
\colhead{$m_{555}$}& 
\colhead{$\sigma_{555}$}& 
\colhead{$m_{814}$}&
\colhead{$\sigma_{814}$}\\
\colhead{\#}& 
\colhead{(J2000.0)}& 
\colhead{(J2000.0)}& 
\colhead{(mag)}& 
\colhead{(mag)}& 
\colhead{(mag)}& 
\colhead{(mag)}
}
\startdata
      1& 01~29~54.828& $-$73~32~31.524&  14.690&   0.004&  13.217&   0.003\\
      2& 01~29~24.581& $-$73~33~16.236&  13.881&   0.002&  14.128&   0.003\\
      3& 01~29~14.220& $-$73~31~53.256&  14.717&   0.003&  13.354&   0.002\\
      4& 01~29~33.310& $-$73~33~44.496&  14.270&   0.002&  14.508&   0.004\\
      5& 01~29~31.051& $-$73~33~43.056&  15.371&   0.004&  15.612&   0.007\\
      6& 01~29~28.301& $-$73~31~57.432&  15.341&   0.004&  15.288&   0.006\\
      7& 01~29~36.485& $-$73~34~12.108&  16.412&   0.007&  14.993&   0.005\\
      8& 01~29~32.837& $-$73~33~39.024&  15.467&   0.004&  15.673&   0.007\\
      9& 01~29~31.992& $-$73~33~30.924&  15.594&   0.005&  15.796&   0.007\\
     10& 01~29~27.768& $-$73~32~59.496&  15.730&   0.005&  15.699&   0.007\\
     11& 01~29~18.286& $-$73~32~33.468&  16.052&   0.006&  15.322&   0.006\\
     12& 01~29~31.094& $-$73~33~43.272&  15.816&   0.005&  16.074&   0.008\\
     13& 01~29~31.411& $-$73~33~42.336&  15.663&   0.005&  15.871&   0.008\\
     14& 01~29~14.249& $-$73~31~53.220&  17.910&   0.020&  16.655&   0.018\\
     15& 01~29~32.942& $-$73~34~10.956&  16.232&   0.006&  16.377&   0.010
\enddata



\tablecomments{Magnitudes are given in the Vega system. Units of right
ascension are hours, minutes, and seconds, and units of declination are
degrees, arcminutes, and arcseconds. Table \ref{t:catalog} is available
in its entirety in the electronic edition of the {\em Astrophysical
Journal}. A portion is shown here for guidance regarding its form and
content.} 


\end{deluxetable}

\begin{figure*}[t!]
\plotone{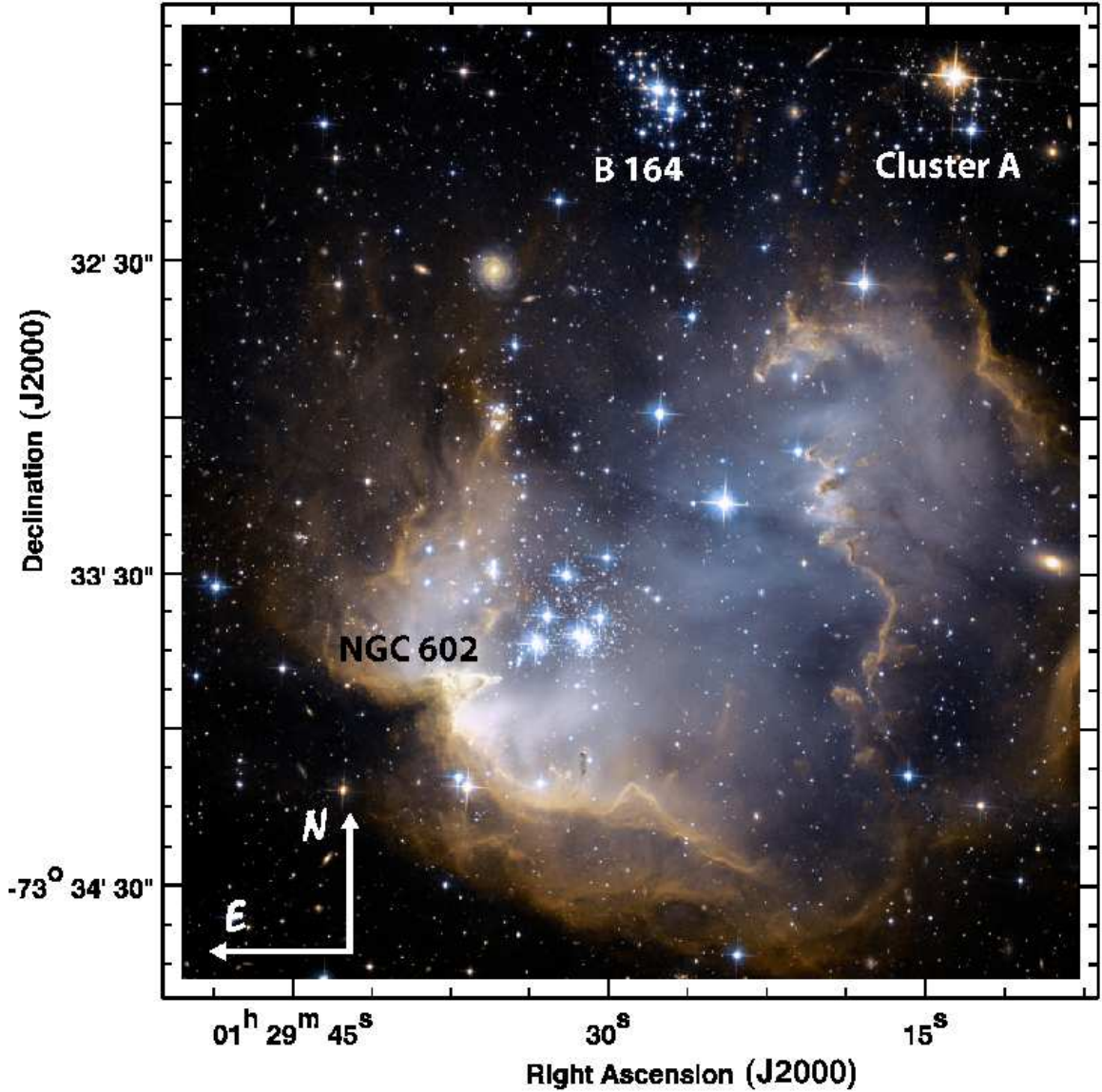}
\caption{Color-composite image of the field around NGC~602, constructed
from observations with HST/ACS in the filters $F555W$, $F814W$ and $F658N$.
The association itself lies at the center of the field, located almost
in the middle of the shell-like emission nebula N~90. Directly north
of the association lies the open cluster B~164 (Bruck 1976), and the
newly discovered ``cluster A'' covers the north-western edge of the field.
All three systems are confirmed as statistically significant stellar
concentrations from star counts shown in Figure~\ref{f:sc}. Credit:
NASA, ESA, and the Hubble Heritage Team (STScI/AURA)-ESA/Hubble
Collaboration.\label{f:rgb}}
\end{figure*}

The investigation by Gouliermis et al. (2005) revealed, for the first
time, that the low-mass stellar content of stellar associations in the
MCs is still in its pre-main sequence (PMS) phase, as is the case of
nearby galactic associations (Brice{\~n}o et al. 2007). Gouliermis et
al. (2006a) discovered in LH~52 about 500 low-mass PMS stars down to
$\sim$ 0.5 M{\solar}, easily distinguishable in the $m_{555}-m_{814}$,
$m_{555}$ Color-Magnitude Diagram (CMD), extending the sub-solar stellar
membership of MCs stellar associations to their PMS populations.
Naturally, these stars should be considered as the best tracers of the
sub-solar IMF in the MCs, but the WFPC2 data, being limited by
completeness, allowed the construction of the PMS IMF of LH~52 only for
the mass range 0.8 - 1.4~M{\solar}. This IMF was found to correspond to
a power-law with a slope of $\Gamma \sim -1.26$. Previous PMS studies in
the MCs focus on the surrounding field of supernova 1987A (e.g. Panagia
et al. 2000), where the sample is limited, and the star-burst of 30
Doradus (e.g. Zinnecker 1998; Sirianni et al. 2000; Brandner et al.
2001), where crowding and high extinction limit the detection of stars
to 1 - 2~M{\solar} (Romaniello et al. 2006). 

\begin{deluxetable}{cccccc}
\tablewidth{0pt}
\tablecaption{Populations Summary in the Identified Stellar Systems
\label{t:roi}}
\tablehead{
\colhead{Name} & 
\colhead{$n_{\rm stars}$} & 
\colhead{$\rho$ (pc$^{-2}$)} &
\colhead{$f_{\rm PMS}$} & 
\colhead{$f_{\rm UMS}$} & 
\colhead{$f_{\rm LMS}$}
}
\startdata
NGC~602   & 855 & 5.6 & {\bf 0.86} & 0.07 & 0.04\\
B~164     & 356 & 2.9 & 0.03       & 0.27 & {\bf 0.62}\\
Cluster~A & 171 & 2.0 & 0.05       & 0.21 & {\bf 0.63}\\
Field     & 230 & 1.1 & 0.16       & 0.08 & {\bf~0.66}
\enddata

\tablecomments{The total stellar numbers of the systems,
$n_{\rm stars}$, are field-subtracted (see \S~4.3.1), and refer to
the area of each system defined with star counts. The total number of
field stars, shown in the last row, refers to the area defined as the
best representative of the field due to its emptiness.  The fractions
$f$ give the number of one stellar species over the total stellar
numbers in every area. The most significant numbers are highlighted. It
is interesting to note that both B~164 and Cluster~A have almost
identical numbers of UMS and LMS stars, and that their LMS fractions are
comparable to that of the field.} 

\end{deluxetable}

Over the last years, stellar associations in the SMC gained increasing
scientific interest, as the investigation of their bright OB stars was
complimented by new studies on their PMS stars (e.g. Nota et al. 2006;
Gouliermis et al. 2006b; Sabbi et al. 2007; Hennekemper et al. 2008).
These studies were facilitated by the unique combination of
high-resolving power with a wide field-of-view provided by the {\em
Advanced Camera for Surveys} (ACS) on-board {\em HST}. In the present
study we deal with the IMF of the young stellar association NGC~602,
related to the bright {\sc H~ii} region LHA~115-N90, or in short N~90
(Henize 1956). The region NGC~602/N~90, being located in the wing of the
SMC, has the advantage of avoiding the densest parts of the galaxy. As a
consequence, the confusion of its stellar content with that of the
general SMC field, which extends in a wide depth along the
line-of-sight, is rather low. McCumber et al. (2005) found in their
investigation on field stars in the SMC wing that after a long period of
constant star formation, there was a burst in activity within the last
1~Gyr. The lack of $\gamma$-ray, X-ray or far-UV sources in the region
of NGC~602/N~90 suggests that this excess in the star formation of the
area was possibly triggered by encounters with the LMC and/or the Milky
Way. Most likely, all major star-forming events in the SMC are initiated
by these tidal interactions. In the first part of our investigation of
the region NGC~602/N~90 (Gouliermis et al. 2007) we present the results
of our photometric study from observations with the {\em Infrared Array
Camera} (IRAC) on-board the {\em Spitzer Space Telescope}. We report the
detection of 22 candidate Young Stellar Objects (YSOs), which are
located at the edge of N~90, suggesting that they are the products of
star formation triggered by the central association NGC~602.

The recent release of the HST/ACS observations of the region of
NGC~602/N~90 offers a unique opportunity for the comprehensive
photometric analysis of both its bright and faint stellar content, and
the construction of the complete IMF throughout the whole observed mass
range. Our photometry, which reaches the limit of $m_{555}$~\gsim~26.5~mag,
providing one of the most complete stellar samples ever collected for
star-forming regions in the SMC, is described in \S~2. We investigate
the morphology of the observed region and the spatial distribution of
the detected stars in \S~3, and in \S~4 we discuss the nature of the
detected stellar species and present the CMDs of the three major stellar
concentrations in the observed field. In \S~4 we also present the
subtraction of the contribution of the field from the observed CMDs, and
we estimate the reddening towards the identified systems. A detailed
description of the stellar members of the association NGC~602 is also
given in \S~4. The Luminosity Functions of the identified stellar
systems are constructed in \S~5. In \S~6 a comprehensive construction of
the IMF of NGC~602, as well as of the two additional star clusters in
the region is presented. Conclusions are given in \S~7.

\section{Observations and Photometry}
\label{s:obs}

The data used in our analysis are obtained with the Wide Field Channel
(WFC) of the ACS within the \hst GO Program 10248 on July 14 (Dataset:
J92F05) and July 18 (Dataset: J92FA6) 2004. Images are taken in the
$F555W$ and $F814W$ filters. A single ACS/WFC pointing covering \ap
3\farcmin4\x3\farcmin4 was centered on NGC~602/N~90. At the distance of
the SMC this field-of-view (FoV) corresponds to an extent of \ap
57\x57~pc. The first visit (J92F05) includes five long exposures taken
with a dithering pattern in order to cover the inter-chip gap of the
camera and one short exposure with a single pointing to cope with the
saturation of the brightest stars. Within the second visit (J92FA6) a
short exposure was also taken in order to remove cosmic rays that were
unintentionally included during the first visit. In Table~\ref{t:obs} we
give a summary of the different data sets used in this investigation. A
color-composite image of the observed field is shown in
Fig.~\ref{f:rgb}.

Photometry was performed with the ACS module of the package
DOLPHOT\footnote{The ACS module of DOLPHOT is an adaptation of the
photometry package HSTphot (Dolphin 2000). It can be downloaded from
{\tt http://purcell.as.arizona.edu/dolphot/}.} (Ver. 1.0), especially
designed for ACS. We followed the photometric process, as it is
described by Gouliermis et al. (2006b). The pipeline-reduced FITS files
were obtained from the \hst Data Archive. We used the package {\em
multidrizzle} (Koekemoer et al. 2002) following the ACS data handbook to
clean the images of residual warm pixels and cosmic rays and to
construct a deep reference drizzled image to be used for the photometry.
All exposures were photometered simultaneously, using the $F814W$
drizzled frame as the position reference. Photometric calibrations and
transformations were made according to Sirianni et al. (2005). Charge
Transfer Efficiency corrections were applied according to ACS~ISR~04-06.
We cleaned our photometric catalog from bad detections using DOLPHOT's
star quality parameters, as described also by Gouliermis et al. (2006b).
The final photometric catalog includes 5,626 stars in total detected in
both filters. The full length of this catalog with the RA, DEC
coordinates (in J2000) and their magnitudes in each filter (in the Vega
system) of these stars is available in electronic form in
Table~\ref{t:catalog}. Astrometric solutions for the detected stars are
derived from the $F814W$ drizzled frame with the use of the application
{\em xy2sky} of the WCSTools package\footnote{The WCSTools package is
available at CfA at {\tt
http://tdc-www.harvard.edu/software/wcstools/}.}.

\begin{figure}[t!]
\epsscale{1.15}
\plotone{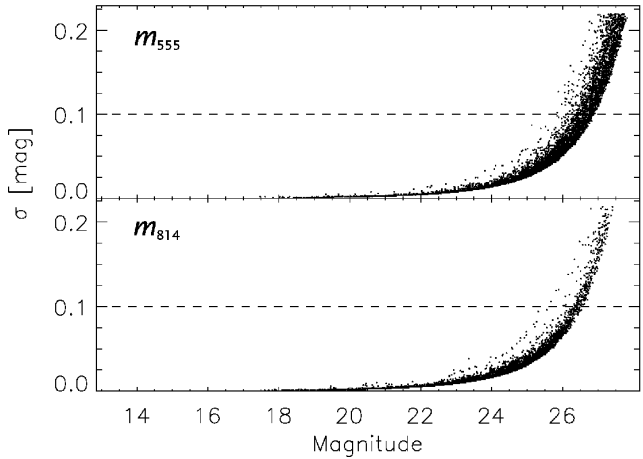}
\caption{Typical photometric uncertainties in both $m_{555}$ and
$m_{814}$ bands as derived by DOLPHOT from all data sets. \label{f:err}}
\end{figure}

\begin{figure}[t!]
\epsscale{1.15}
\plotone{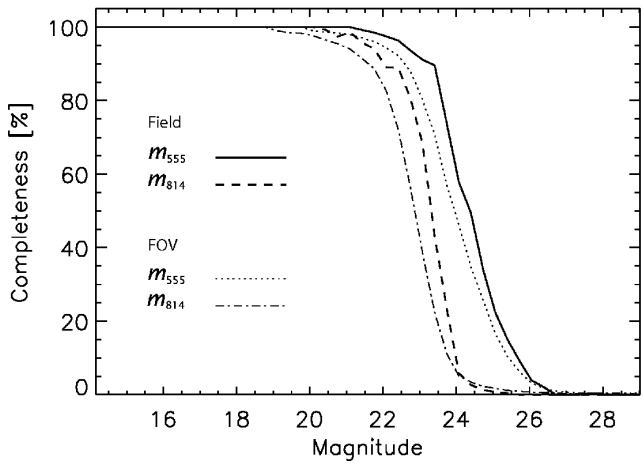}
\caption{Completeness for all stars in the observed FoV and in the area
selected as the most representative of the local field the SMC. The
field being less contaminated by the nebula and less crowded than the
association NGC~602 exhibits a higher completeness than the overall FoV.
\label{f:cp}}
\end{figure}

Typical uncertainties of our photometry as a function of the magnitude
for both filters are given in Fig.~\ref{f:err}. The completeness of the
data was evaluated by artificial star experiments with the use of lists
of almost 400,000 artificial stars created with the utility {\em
acsfakelist} of DOLPHOT. The completeness is found to be spatially
variable, depending on the crowding of each region. This is shown in
Fig.~\ref{f:cp}, where the completeness factors for the whole observed
field, as well as of the area selected as the most representative of the
background field, due to its emptiness, are plotted for both filters.
The area, which from here on we refer to as simply the ``field'', is
selected as the most {\em empty area} of the western part of the
observed FoV.  This selection was based on our recent results from {\em
Spitzer}/IRAC, which show that this part of NGC~602/N~90 is the less
contaminated by dust emission (Gouliermis et al. 2007).

\section{Morphology of the region of NGC~602/N~90}

The prominent bright stellar concentration almost at the center of the
observed ACS/WFC field, shown in Fig.~\ref{f:rgb}, is the association
NGC~602. The bright emission nebula, N~90, can also be seen in the
image, because of the H$\alpha$ emission, included in the $F658N$ filter
(centered at 658.4~nm). A small contribution from [N{\sc ii}] is also
included in this filter. The association is surrounded by a ring-like
feature highlighted by the gas emission, and ribbon-like structures
pointing towards the center can be seen to the southeastern and
northwestern of the ring. These structures have been identified by
Gouliermis et al. (2007) as the SMC analogs of the ``Pillars of
creation'' originally detected in the Galactic {\sc H~ii} region M~16
(Hester et al. 1996). They clearly indicate ongoing star formation on
the rim of the ring-like structure, probably driven by the stellar winds
and UV radiation of the brightest stars in NGC~602. 

The stellar association is not the only stellar concentration inside the
observed FoV. North of the association lies the open cluster B~164
(Bruck 1976) which is also known as NGC~602b (Westerlund 1964). Next to
B~164 and to the west lies a third stellar concentration. It shows to be
a cluster, which does not belong to any catalog of known objects in this
region. From here on we refer to this cluster as ``Cluster~A''. All
three stellar systems are easily revealed by the spatial distribution of
all stars detected with our photometry in the region, as discussed in
the next section.

\begin{figure}[t]
\epsscale{1.15}
\plotone{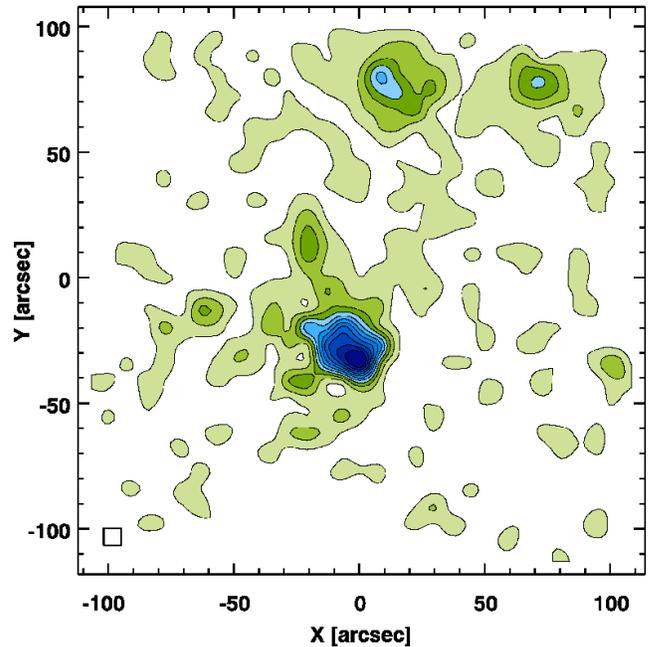}
\caption{Isodensity contour map for all stars with photometric errors
$\sigma_V\leq$~0.1~mag using grid elements of size 7\arcsec,
shown as the small box at the bottom-left corner of the map. The first
isopleth corresponds to the mean background density. The next isopleths
are plotted in steps of the standard deviation $\sigma$. The bluish
contours correspond to stellar density equal or higher than the
3$\sigma$ level. All three stellar systems in the region are revealed as
significant stellar concentrations.\label{f:sc}}
\end{figure}



\subsection{Spatial Distribution of the Stars in NGC~602/N~90\label{s:iso}}

In order to reveal the spatial distribution of the detected stars and
the morphology of any stellar concentrations in the observed region, we
performed star counts on our photometric catalog (a sample of this
catalog is shown in Table~\ref{t:catalog}). Stars with photometric
uncertainties $\sigma_{V}\geq$~0.1~mag were not
considered. The constructed isodensity contour map is shown in
Fig.~\ref{f:sc}. The coordinate system used for this map, as well as
the ones shown later is the Cartesian (X,Y) system of the pixel
coordinates of WFC, transformed to seconds of arc, in respect to the
central pixel of the FoV.

Star counts were performed in a quadrilateral grid divided in elements
with sizes 140~$\times$~140 WFC~pix$^2$ each, which correspond to
7\arcsec\x7\arcsec\ (or $\approx$~2~pc~$\times$~2~pc at the distance of
the SMC). This size was selected as the optimum to map any fine
structure, which might exist within the clusters. The selection of
larger size for the grid elements would reveal a smoother version of the
map of Fig. ~\ref{f:sc}. The first isopleth in the contour map indicates
the mean background density, whereas the subsequent levels are drawn in
steps of 1$\sigma$, where $\sigma$ is the standard deviation of the
background density. All contours with density equal or higher than the
3$\sigma$ level, which is chosen as the density threshold of the
statistically significant stellar concentrations are drawn with bluish
colors. All three star clusters in the region are revealed in this map
with NGC~602 being the dominant stellar system, seen almost at its
center. The density peaks for both B~164 and Cluster~A do exceed the
3$\sigma$ limit as well. Smaller compact density peaks can be seen in
the map below the 3$\sigma$ limit. They are typical density
fluctuations, due to the small size selected for the grid elements. 

\section{Stellar Populations in the Region of NGC~602/N~90}

\subsection{Color-Magnitude Diagram\label{s:cmd}}

\begin{figure}[t!]
\epsscale{1.15}
\plotone{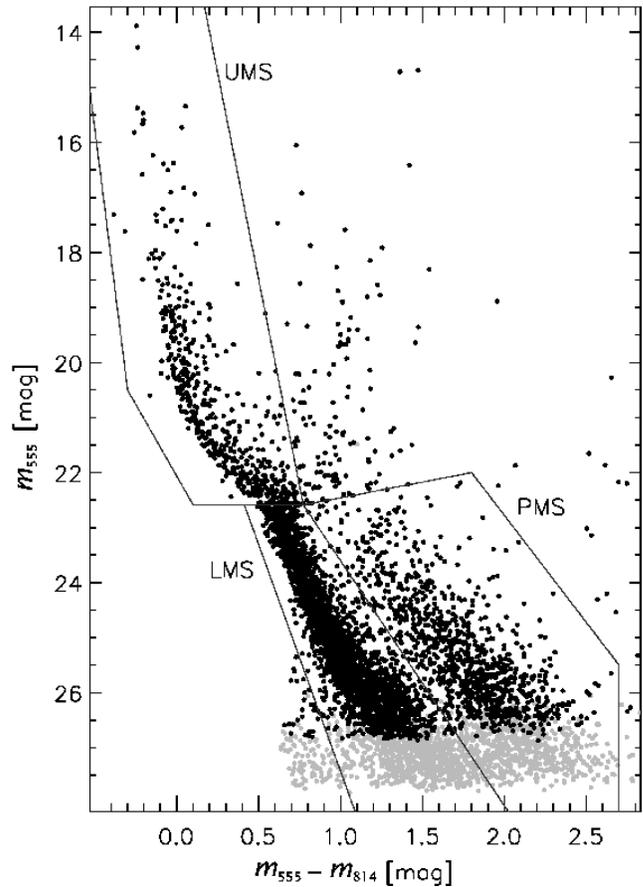}
\caption{The $m_{555}-m_{814}$, $m_{555}$ CMD of all stars detected in
the region of NGC~602/N~90. Stars with photometric uncertainties larger
than 0.1~mag (Fig.~\ref{f:err}) are shown as grey dots. The drawn lines
indicate the tentative limits, which separate the different stellar
species included in the catalog of detected stars: The Upper Main
Sequence (UMS), the Lower Main Sequence (LMS), and the Pre-Main Sequence
(PMS) stars.\label{f:cmd}}
\end{figure}

The $m_{555}-m_{814}$, $m_{555}$ color-magnitude diagram (CMD) of all
stars in our photometric catalog is shown in Fig.~\ref{f:cmd}, which
indicates that our photometry provides accurate magnitudes (with
$\sigma_{555}$~\lsim~0.1~mag) for stars with $m_{555}$~\lsim~26.5~mag.
Different groups of stellar types can easily be distinguished in this
CMD.  There is a prominent well populated main sequence (MS), which
extends from the detection limit up to $m_{555}\approx$~14~mag, and a
rich population of pre-main sequence (PMS) stars located on the red part
of the MS. We select regions of the CMD, each covering different types
of stars.  The tentative limits of these regions are also shown in
Fig.~\ref{f:cmd}. We split the MS into its upper (brighter) and lower
(fainter) parts (UMS and LMS respectively). According to the ZAMS
isochrone from the model grid designed by Girardi et al. (2002) for the
ACS filter system, the limit between UMS and LMS stars of
$m_{555}=22.6$~mag corresponds to a stellar mass of
$M\approx$~1.2~M{\solar}. 

The brighter limit of the PMS region in the CMD is being chosen around
$m_{555}\approx$~22.3~mag. Probably, a brighter limit would allow us to
include the brighter PMS stars and the transition region between PMS and
MS (turn-on), but this part of the CMD is contaminated by the old
stellar population of the field of the galaxy (e.g. McCumber et al.
2005), as seen by its loose red giant branch (RGB). The low numbers of
widely scattered stars in the RGB part of the CMD and the absence of any
clearly detectable turn-off indicates that only a small fraction of the
evolved field stars can be seen in the observed region, probably due to
extinction by the dust. Indeed, {\em Spitzer}/IRAC observations of
NGC~602/N~90 have shown in the 8~\micron\ band strong dust emission from
the area, which surrounds the central ``hole'' in the {\sc H~ii} region,
where NGC~602 is located (Gouliermis et al. 2007).

\begin{figure*}[t!]
\epsscale{1.1}
\plottwo{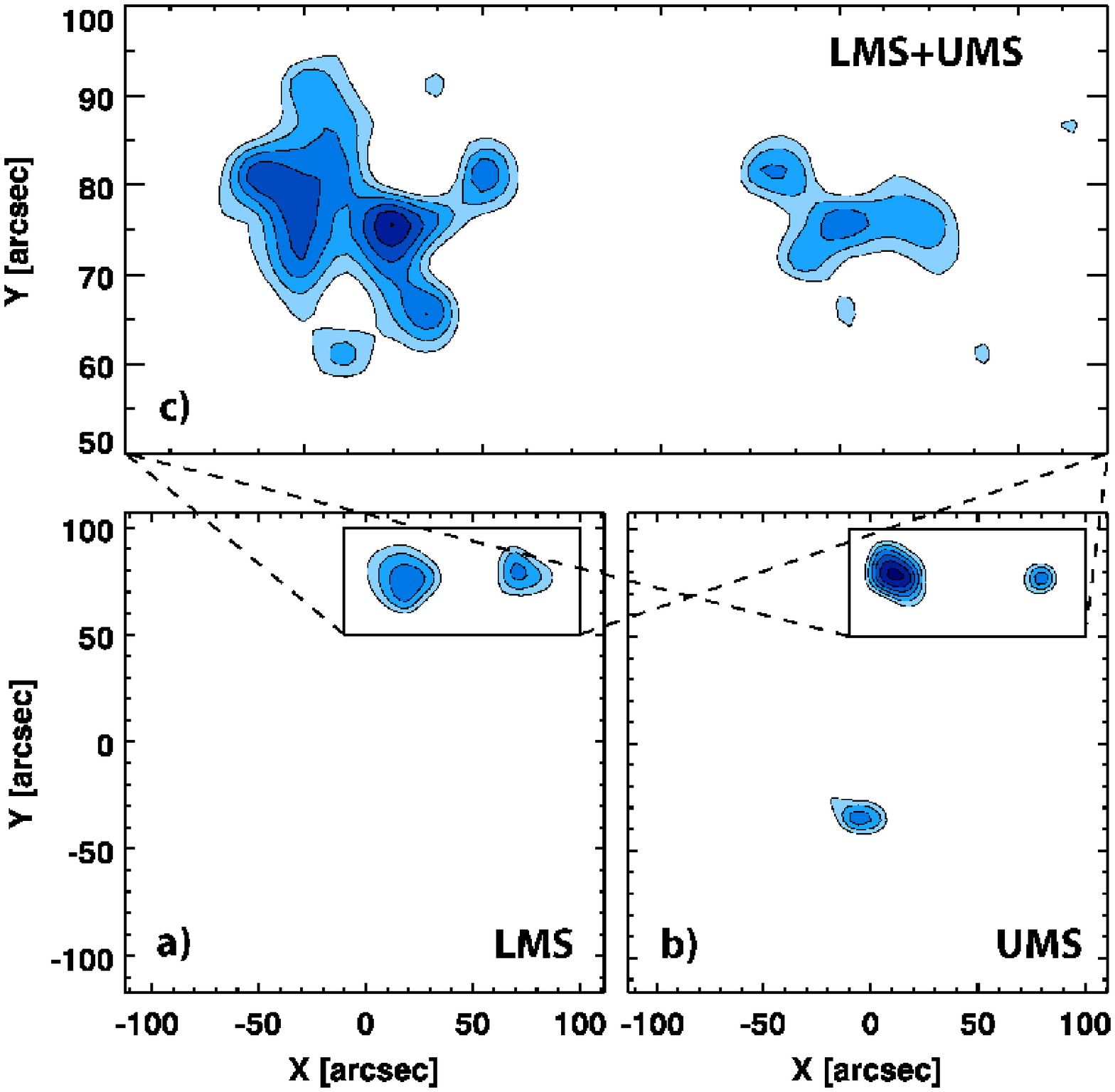}{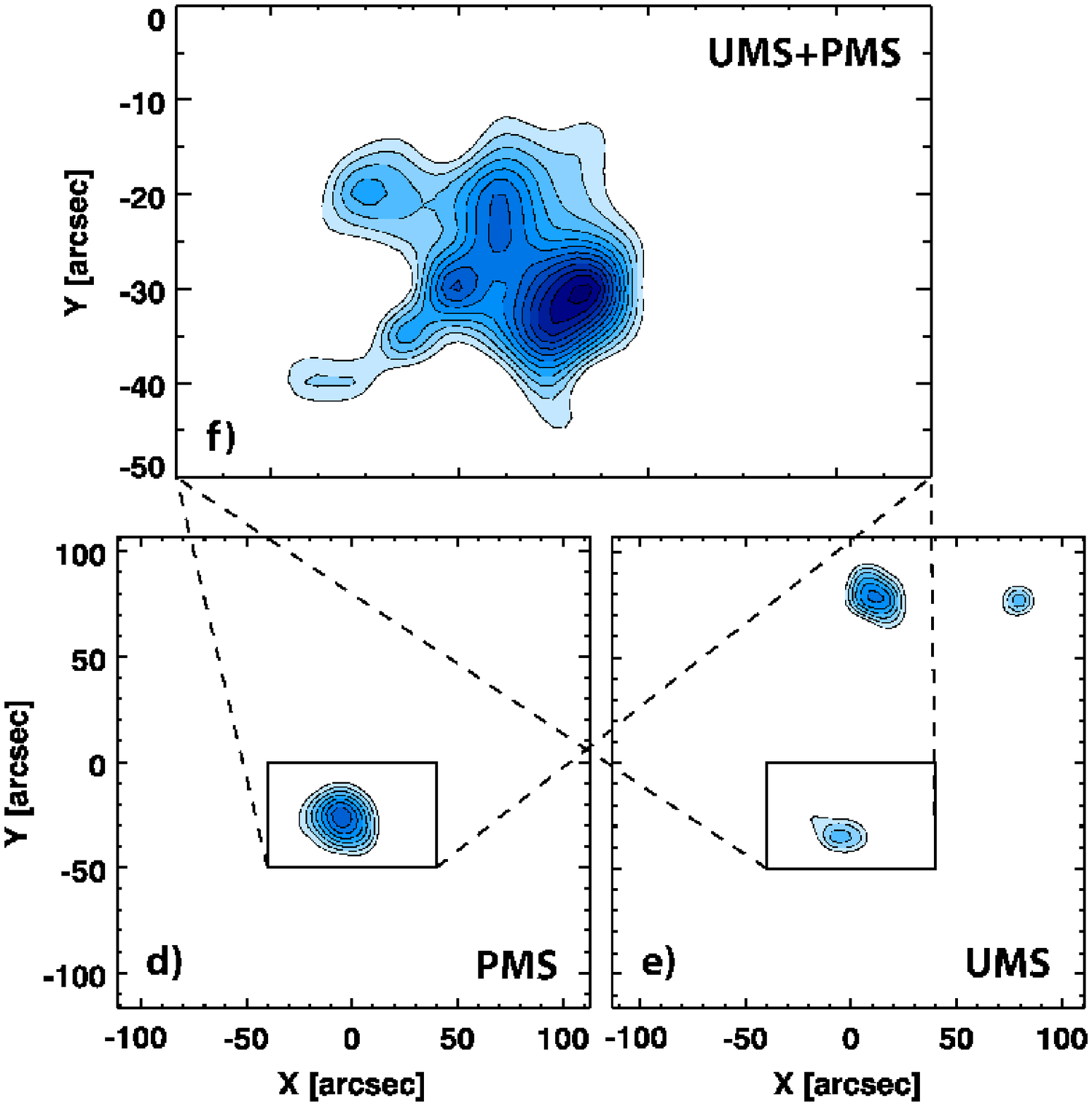}
\caption{Isodensity contour maps for the region of NGC~602/N~90 from
star counts of different stellar populations (LMS, UMS, PMS; see \S
4.1), as they are revealed in the observed CMD. The maps of the whole
region (bottom panels) have a grid resolution of 12\arcsec, while the
enlarged ones (top panels) of 5\arcsec. In all maps the first drawn
density isopleth corresponds to the 3$\sigma$ significance.\label{f:c}}
\end{figure*}

\subsection{Stellar Populations in the  Systems of the Region\label{s:sys}}

Isodensity contour maps from counts of the three most important stellar
types in the CMD of Fig.~\ref{f:cmd} exhibit a strong spatial
distinction between UMS, LMS and PMS stars. This is shown in the bottom
panels of Fig.~\ref{f:c} (Figs.~\ref{f:c}a, \ref{f:c}b, \ref{f:c}d and
\ref{f:c}e). The contour maps of this figure are constructed with the
same method as the map of Fig.~\ref{f:sc} (see \S~3.1), and the lowest
density isopleth shown corresponds to 3$\sigma$ above the local
background density. These maps reveal again all three important stellar
systems in the region (NGC~602, B~164 and Cluster~A), but they
demonstrate that these systems are dominated by different types of
stars. 

All three systems are detected by star counts of UMS stars. Although
B~164 and Cluster~A are found to contain significant samples of LMS
stars, they are not revealed at all in the map of the PMS stars
(Fig.~\ref{f:c}d). On the other hand, the faint stellar populations of
the association NGC~602 are certainly PMS and not LMS stars. This
demonstrates that both B~164 and Cluster~A seem to be evolved small
clusters, but NGC~602 is a younger stellar association, with a
significant number of PMS stellar members. Consequently, in order to
reveal the extent of each system based on its dominant stellar
populations we construct their contour maps from star counts of these
populations alone, shown in the top panels of Fig.~\ref{f:c}. The map
constructed for B~164 and Cluster~A from the UMS and LMS stars is shown
in Fig.~\ref{f:c}c, while the one for NGC~602 (from counts of the UMS
and PMS stars) is shown on Fig.~\ref{f:c}f. We define the boundaries of
each stellar system from the 3$\sigma$ isopleth in these maps. The size
of each system is defined by a rectangle, which contains this isopleth.

We show in Table~\ref{t:roi} the numbers of stars counted within the
measured limits of each system (after the field contribution subtracted,
see 4.3.1), the surface density and the corresponding fractions of PMS,
UMS, and LMS stars included in these limits. These fractions show a
clear concentration of PMS stars in NGC~602, while Cluster~A and B~164
clearly host the largest amount of LMS stars. Although, both Cluster~A
and B~164 show to host larger numbers of UMS stars, it is NGC~602 that
hosts the brightest MS stars in the region, as it will be shown later
from the CMDs of the systems. Considering that the limits for the UMS
stars in the CMD have been selected to include stars down to
$m_{555}\approx$~22.5~mag, the larger fraction of UMS stars in Cluster~A
and B~164 is due to their fainter UMS members.

\subsection{The Star Clusters B~164 and Cluster~A}
\label{s:pop}

According to our selection of different stellar types in the observed
CMD (\S~4.1), about 55\% of the stars in our catalog are LMS stars. 
Fig.~\ref{f:c}a shows that these stars are mostly concentrated in B~164
and Cluster~A, but in a rather loose manner (there are only three
isopleths above the local background). On the other hand, the
concentration of UMS stars in these clusters shows a peak at 9$\sigma$
above the local background in B~164 (Fig.~\ref{f:c}b). The detailed
contour map of the area around these clusters, shown in Fig.~\ref{f:c}c,
shows the prominent concentration of both UMS and LMS stars in both of
them (and mostly in B~164) indicating their evolved nature. The absence
of PMS stars in both clusters as seen in the map of Fig.~\ref{f:c}d,
certainly affects their individual CMDs. In order to construct these
CMDs we consider the contribution of the stellar population of the
general field of SMC in the region, as it is observed within the
observed ACS pointing. We apply a Monte Carlo method for the subtraction
of this contribution from the CMD observed in the specific area of each
cluster.

\subsubsection{Subtraction of Contaminating Field Stars}

The typical procedure for the field subtraction of a CMD with the use of
the Monte Carlo method divides the CMD in a grid and performs a
comparison between the numbers of stars found in each grid element of
the observed CMD of the cluster and the one of the region selected as
representative of the general field. Although this method is efficient
for compact populous star clusters, it might produce unwanted artifacts
for smaller not compact clusters or associations. Therefore, our Monte
Carlo method for the field subtraction of the observed CMDs of the
clusters slightly differs from the typical procedure. Our
field-subtraction procedure does not use a predefined grid on the whole
extent of the CMDs, but an individual subregion of the CMD is considered
around every single star observed in the area of the cluster.  Each such
subregion has the shape of an ellipse with semi-axes $\Delta(V-I)=
0.15$~mag and $\Delta V=0.5$~mag.

The corresponding numbers of field and cluster stars inside each of
these ellipses ($n_{\rm F}$ for the field and $n_{\rm C}$ for the
cluster) are counted, and corrected for incompleteness according to the
completeness factors estimated per magnitude range for each of the
considered areas. The number of field stars per ellipse expected to be
present in the CMD of the cluster, $n_{\rm F}^*$, is given by the number
of field stars, $n_{\rm F}$, normalized to the surface of the area of
the cluster. The probability that the central star of each ellipse is
actually a field star is given by $n_{\rm F}^*/n_{\rm C}$. If e.g.
$n_{\rm F}^*=0$ then all the stars inside the ellipse are cluster stars
and the probability is 0, while if $n_{\rm F}^*/n_{\rm C}\geq 1$ then we
can assume that this star is actually a field star. For intermediate
values, $0 < n_{\rm F}^*/n_{\rm C} < 1$, it is randomly decided whether
to account this star for a field star or not.

\begin{figure}[tb!]
\epsscale{1.2}
\plotone{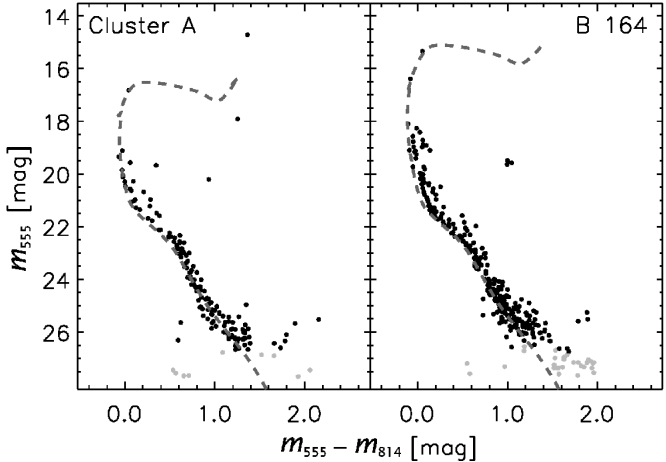}
\caption{The $m_{555}-m_{814}$, $m_{555}$ CMD of the stars in the
regions of Cluster~A and B~164, after the contribution of the field has
been statistically subtracted with the Monte Carlo method. Their
prominent MS shows that both clusters are most probably evolved open
clusters. The overplotted isochrones from the grid of evolutionary
models by Girardi et al. (2002) represent the older limits in the age
estimation of each cluster.
\label{f:cmdab}}
\end{figure}

We applied this method to decontaminate the CMD of the area of each
system from the contribution of the field. The field-subtracted CMDs of
both Cluster A and B~164 are shown in Fig.~\ref{f:cmdab}. Their most
remarkable feature is the clear MS, which extends from the detection
limit up to the brightest stars in the clusters, with magnitudes
$m_{555}\approx$~16.8~mag for Cluster~A and $m_{555}\approx$~15.5~mag
for B~164. An older limit for the age of both clusters can be estimated
based on their bright MS stars, with isochrone fitting. We used the
models by Girardi et al. (2002), especially designed for the filter
system of ACS, and we found an age of $\approx$~160~Myr for Cluster~A,
while B~164 is found be not older than $\approx$~80~Myr. It should be
noted that if the only bright red star seen in the CMD of Cluster~A is
also considered, an age of $\tau \sim$~80~Myr may also represent this
cluster. If indeed the two clusters have similar ages, and considering
that they are very close to each other, one may speculate that both
clusters form a double system, product of the same star formation event.
However, the available data are not sufficient to support or dismiss
this hypothesis.

In Fig.~\ref{f:cmdab} both CMDs are shown with the corresponding
isochrones overplotted. In these CMDs, as well as in all CMDs shown here
we assume a distance modulus $\mu_{\rm 0} \simeq 18.9 \pm 0.1$~mag
(Dolphin et al. 2001). The largest observed stellar masses for MS stars
in each cluster can be estimated from the evolutionary models
considered.  These masses are $\sim$~4.3~M{\solar} for Cluster~A and
$\sim$~5.7~M{\solar} for B~164.

\begin{figure}[t!]
\epsscale{1.2}
\plotone{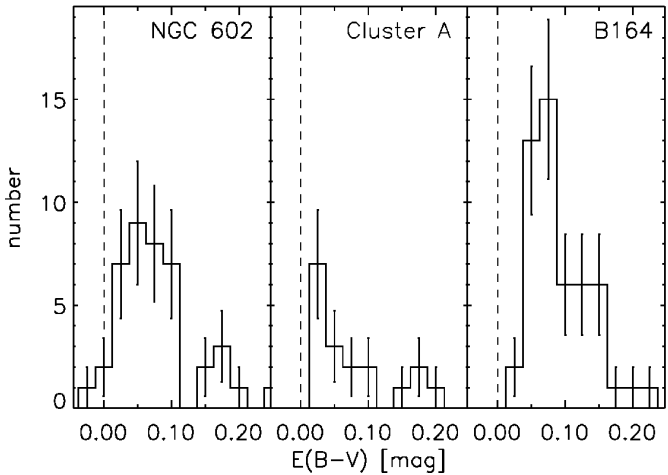}
\caption{The $E(B-V)$ color excess distributions derived from the bright
MS stars with $m_{555}< 21.25$~mag in NGC~602 and the clusters B~164
and Cluster~A. The peaks of the distributions were estimated assuming
that they correspond to a Gaussian distribution. The obtained values are
for NGC~602 $E(B-V)=0.06\pm0.02$~mag, for Cluster~A
$E(B-V)=0.04\pm0.02$~mag, and for B~164 $E(B-V)=0.07\pm0.02$~mag. These
values show that all three clusters suffer from rather low optical
extinction. \label{f:red}}
\end{figure}

\subsubsection{Interstellar Reddening\label{s:red}}

For the isochrones plotted in Fig.~\ref{f:cmdab} no specific
interstellar reddening was considered. The comparison of the loci of the
individual UMS stars to the models allowed us an accurate determination
of the reddening within the area of each cluster. The corresponding
distributions of the stars in Cluster~A and B~164 according to their
reddening are shown in Fig.~\ref{f:red} (the distribution for NGC~602 is
also shown for comparison; see \S~4.4). These plots show that a mean
color excess of $E(B-V)= 0.04 \pm 0.02$~mag can be considered for
Cluster~A and $E(B-V)= 0.07 \pm 0.02$~mag for B~164. These values, as
well as the one for NGC~602, are in excellent agreement with previous
reddening measurements, which vary between $E(B-V)\simeq 0.03$ and
$0.07$~mag for clusters located in the Wing of the SMC (Piatti et al.
2007), and between $E(B-V)\simeq 0.05$ and $0.09$~mag for southern
fields of the galaxy (No{\"e}l et al. 2007). They are also in good
agreement with previous measurements on SMC intermediate-age clusters
(e.g. Alcaino et al. 2003; Hunter et al. 2003; Rochau et al. 2007). The
reddening values found for Cluster~A and B~164 correspond to an
extinction of $A_V\approx$~0.1~mag and $A_V\approx$~0.2~mag respectively
assuming an extinction law $A_V=R_V\;E(B-V)$ with $R_V=3.1$ (e.g.
Koornneef 1983). 

\subsection{The Stellar Association NGC~602}
\label{s:popngc}

As mentioned earlier (\S~4.2) NGC~602 does not host significant numbers
of LMS stars (Fig.~\ref{f:c}a), but rather a large concentration of PMS
stars (Fig.~\ref{f:c}e). The system is also characterized by a prominent
UMS stellar population (Fig.~\ref{f:c}d). Fig.~\ref{f:c}f shows in more
detail the morphology of the system based on the spatial distribution of
both UMS and PMS stars. This map shows clear signatures of
sub-clustering within the system (as is the case for Cluster~A and
B~164, shown in Fig.~\ref{f:c}c), which is demonstrated by the clumpy
behavior of the density in the map. Although, this phenomenon could be
due to the selection of a finer grid resolution in the star counts (with
a grid element size of $\approx$~5\farcs0 instead of 12\farcs0, which
was used for the larger maps), from the positions of the UMS stars we
verified that they are indeed concentrated in smaller subregions within
NGC~602, the most prominent being the one located to the south-east
(bottom-right) of the map in Fig.~\ref{f:c}f. This subgroup has also an
over-density of PMS stars located with an offset of \ap5\arcsec\ in
respect to the density peak of the UMS stars. A second important
subgroup of UMS stars is located \ap 10\arcsec\ east of the first, and
as shown in Fig.~\ref{f:c}f it is somewhat elongated.

\begin{figure}[t!]
\epsscale{1.2}
\plotone{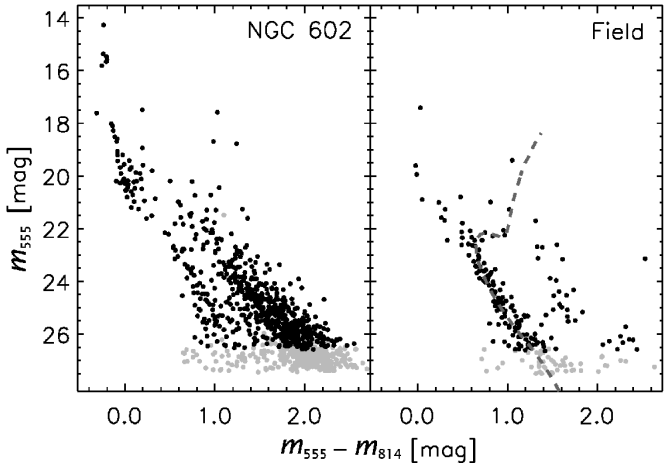}
\caption{The $m_{555}-m_{814}$, $m_{555}$ CMD of the stars in the area
of the association NGC~602 (without any subtraction of the contribution
of the field), and of the area selected as the most representative of
the general background field of the SMC. In the latter, features which
are characteristic of the SMC wing stellar population, such as a turnoff
at $\approx$~22~mag (McCumber et al. 2005), corresponding to the
overplotted isochrone of age $\sim$~1~Gyr, can be distinguished.
\label{f:cmdnf}}
\end{figure}

\begin{figure}[t!]
\epsscale{1.15}
\plotone{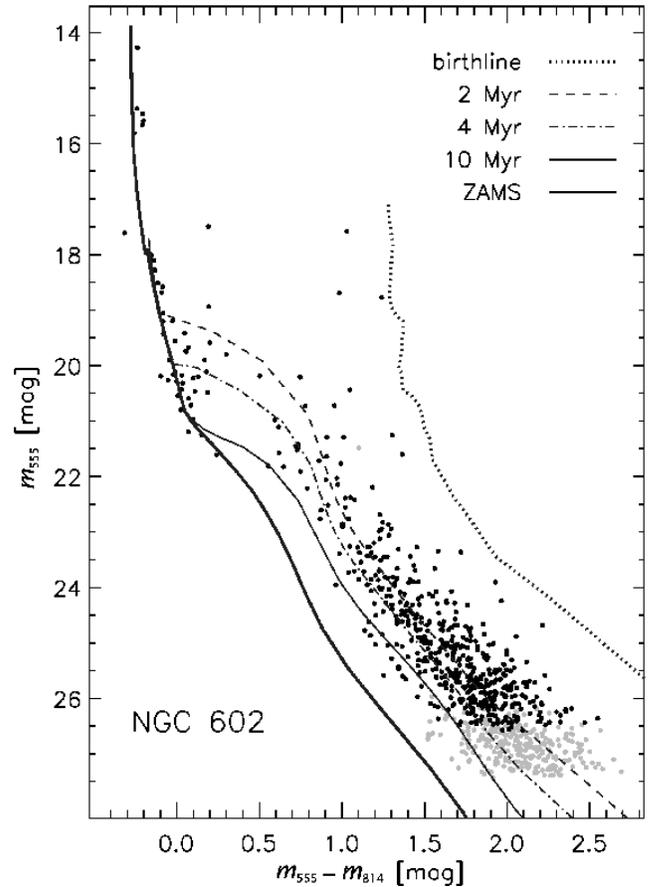}
\caption{The $m_{555}-m_{814}$, $m_{555}$ CMD of the stars in NGC~602,
after the contribution of the field has been statistically subtracted
with the Monte Carlo method. A clear upper MS (down to $m_{555}\simeq
21.5$~mag), as well as a sequence of faint PMS stars (with $m_{555}\gsim
21.5$~mag) can be easily distinguished to represent the stellar
populations of the association.  The loci of the PMS stars show a
prominent broadening.  This makes the estimation of the age of these
stars rather difficult. A discussion on the origins of such a spread in
the positions of PMS stars in the CMD is presented in \S~\ref{s:popngc}.
Characteristic isochrones from the grid of PMS models by Siess et al.
(2000) are overplotted. The birth-line and the ZAMS from the same grid
of models are also plotted. \label{f:cmdngc}}
\end{figure}

The CMD of the region of NGC~602 is shown in Fig.~\ref{f:cmdnf} in
comparison to the CMD of the region of the field. NGC~602 shows a more
well populated and brighter UMS than Cluster~A and B~164, clearly
suggesting its youthfulness. We identified the 10 brightest stars of
NGC~602 as early-type with spectral types between O5.5 and B0.5. These
``photometric spectral types'' are consistent ($\pm 0.5$ spectral type)
with those previously derived by Hutchings et al. (1991) from optical
and UV spectroscopy. Moreover, the most prominent feature in its CMD is
the large number of stars in the PMS part, which is completely empty in
the CMDs of both Cluster~A and B~164 and of the field\footnote{The
selection of the region of the field could not be made without including
some PMS stars. The reason is that no control field was observed away
from NGC~602 and therefore we selected the most ``empty'' and remote
part of the observed FoV as the best representative of the field. The
inclusion of PMS stars in the field may lead to the underestimation of
the the number of PMS stars in NGC~602. However, as derived from the
numbers of Table 3, this cannot affect more than 5\% of the total PMS
members of NGC~602.}. For the subtraction of the contribution of the
field population in the CMD of NGC~602 we applied the same Monte Carlo
method (described in \S~4.3.1) as for Cluster~A and B~164. The derived
``clean'' CMD of the stars of NGC~602 alone, shown in
Fig.~\ref{f:cmdngc}, clearly indicates that all the observed LMS stars
belong to the field and not the association. The distribution of
interstellar reddening for the UMS stars of NGC~602 (Fig.~\ref{f:red})
also shows very low extinction. The mean color excess obtained assuming
a Gaussian distribution is $E(B-V)=0.06\pm0.02$~mag.

In the CMD of Fig.~\ref{f:cmdngc} it can be seen that the positions of
the PMS stars show a spread as if they belong to groups of different
ages. The observed broadening of the loci of the PMS stars in the CMD
implies a wide span in ages of over 10 Myr. Three indicative isochrones
of ages 2, 4 and 10~Myr are overplotted in Fig.~\ref{f:cmdngc} to
demonstrate this spread. If the CMD broadening of the PMS stars is
produced by an age-spread or not is very important in understanding the
star formation history of the region. The existence of an age-spread
would suggest that these stars {\em did not} originate from a single
star formation event, but they were formed over a longer period of time.
On the other hand if all PMS stars are the product of a single star
formation event, then there should be other factors that affect their
CMD positions.

The importance of the age of PMS stars lies on the fact that these stars
are the live records of star formation that took place in a star forming
region within the last \lsim~30~Myr (e.g. Brice{\~n}o et al.\ 2007). 
Consequently, if we assume that different age-groups of PMS stars can be
easily distinguished by their locations in the CMD, they should be the
best tracers of any sequential star formation event that might have
taken place within their host stellar systems. The age-spread of PMS
stars in young stellar systems of our Galaxy has been discussed by e.g.
Palla (2005), who notes that the star formation history of the ONC
extends long in the past, although at a reduced rate, and by Preibisch
\& Zinnecker (2007), who found no evidence for an age dispersion in
Upper Scorpius OB association, although small age spreads of
$\approx$~12~Myr cannot be excluded. 

\begin{figure}[t]
\epsscale{1.15}
\plotone{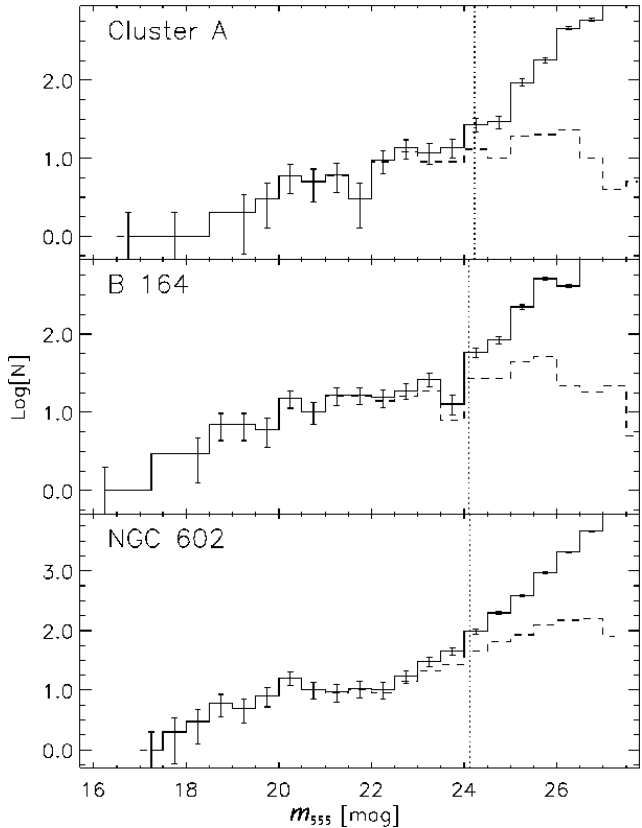}
\caption{The $m_{555}$-luminosity functions (LFs) for all three stellar
systems.  The limit of the 50\% completeness is indicated in each LF by
the dotted lines. The measured numbers of stars are shown with the
dashed lines, while the thick lines show the LFs after the stellar
numbers have been corrected for incompleteness. In all shown LFs the
contribution of field stars has been statistically
subtracted.\label{f:lum}}
\end{figure}

However, considering that PMS stars have a rather peculiar behavior, as
there are several effects that change their position in the CMD, the
retrieval of any signature of age-spread in a PMS population with
single-epoch photometry alone is not a trivial task. Hennekemper et al.
(2008) using the same observational material of another SMC association,
NGC~346, showed that there are various physical properties of the PMS
stars of the system, which affect their loci in the CMD. This produces a
broadening, which could be misinterpreted as an age-spread. These
properties, typical for T~Tauri stars of our Galaxy, are unresolved
binarity, variability and differential extinction. Although it is most
certain that these characteristics are responsible for most of the
observed CMD broadening of the PMS stars (Hennekemper et al. 2008), one
cannot exclude the possibility that there might be also a true
age-spread among these stars, hidden in the CMD of NGC~602. We currently
explore this possibility in another paper by modeling the observed CMD
with the construction of synthetic CMDs for NGC~602 (M. Schmalzl et al.,
in preparation).

\section{Luminosity Functions}
\label{s:lum}

We constructed the luminosity function (LF) of all three stellar systems
by counting the stars in magnitude bins of 0.5~mag in both $m_{555}$ and
$m_{814}$.  The $m_{555}$-LFs of the systems are shown in
Fig.~\ref{f:lum} for the field subtracted stellar samples shown in
Figs.~\ref{f:cmdab} and \ref{f:cmdngc}. The LFs of the counted numbers
of stars are shown with dashed lines, while the LFs shown with thick
lines are corrected for incompleteness. The dotted vertical line for
every plot marks the 50\% completeness limit for each system. The LFs of
both Cluster~A and B~164 are very similar, and they both suffer from
lower number statistics than the LF of NGC~602 shown in the bottom panel
of Fig.~\ref{f:lum}. In the latter one can identify a small dip at
$m_{555}\approx$~22~mag. This feature has been theoretically predicted as a
direct consequence of the stellar evolution of young star clusters,
since the stars at $m_{555}\approx$~21~mag gain brightness very fast,
depleting thus this part of the CMD (Stahler \& Fletcher 1991; Fletcher
\& Stahler 1994).

\section{MASS FUNCTION}

\subsection{Mass Spectrum}
\label{s:msintro}

The stellar Initial Mass Function (IMF), $\xi(\log M)$, is the
distribution of the stars in a stellar system according to their masses,
by the time of their formation. The IMF gives the number of stars of the
system per unit logarithmic (base 10) mass, $d\log{M}$, per unit area. 
There are several functional forms proposed to represent the IMF (see
Kroupa 2002 for a review), but its intermediate- and high-mass part
(down to $M \simeq 1$~M{\solar}) is generally characterized by a
single-power law of the form $\xi(\log M)\sim M^\Gamma$. Typical
logarithmic derivative is $\Gamma \approx -1.35$, as estimated for stars
in the solar neighborhood with masses 0.4~\lsim~$M$/M{\solar}~\lsim~10
(Salpeter 1955). If the stars are counted in unit mass intervals, $dM$,
per unit area, then their Initial Mass Spectrum (IMS), $f(M)$, is
constructed.  IMS is generally described also by a power law of the form
$f(M)\sim M^\alpha$. The IMS and IMF are connected by the relation
(Miller \& Scalo 1979): \begin{equation} f(M)=\xi(\log M)\frac{{\rm
d}\log{M}}{{\rm d}M} \end{equation} Consequently, the relation for their
indexes (slopes) is: \begin{equation}
	\alpha=\Gamma-1
	\label{e:ims2imf}
\end{equation}

For the construction of the IMF or IMS a mass luminosity (ML) relation
based on isochrone models is required for the translation of the
observed stellar luminosities to masses. However, as shown in \S~4.4,
the loci of the PMS stars in the CMD of NGC~602 are well spread,
producing a broadening which covers a wide range of isochrones
(Fig.~\ref{f:cmdngc}). Since the observed CMD broadening of PMS stars is
not solely due to age differences, but also due to the characteristics
of these stars (Hennekemper et al. 2008), a construction of a unique ML
relation is rather impossible. Specifically, the observed optical
magnitudes of young PMS stars are affected by several factors,
especially the extinction/emission produced by the accretion disk and/or
envelope (e.g. Herbst et al. 2002; Sherry et al. 2004; Brice{\~n}o et
al. 2005). At ages less than $10^7$~yr all low-mass PMS stars are
variable, and the strength of this photometric variability depends on
the star's magnetic activity, the amount of circumstellar matter, and
the accretion rate. Classical T~Tauri stars are actively accreting gas
from circumstellar disks, producing irregular photometric variability
with amplitudes up to 3 mag in the $V$ band (Herbst et al. 1994).
Naturally, such variations affect the estimation of the luminosities and
the effective temperatures for these PMS stars, as well as of their
individual circumstellar and differential reddening. Therefore, any
attempt to fit isochrone models to the observed CMD and to determine the
ages to these stars will be biased\footnote{See Hennekemper et al.
(2008; \S~3) for a discussion on the limitations in the age estimation
for the PMS stars in the SMC association NGC~346.}.

\begin{figure}[t]
\epsscale{1.15}
\plotone{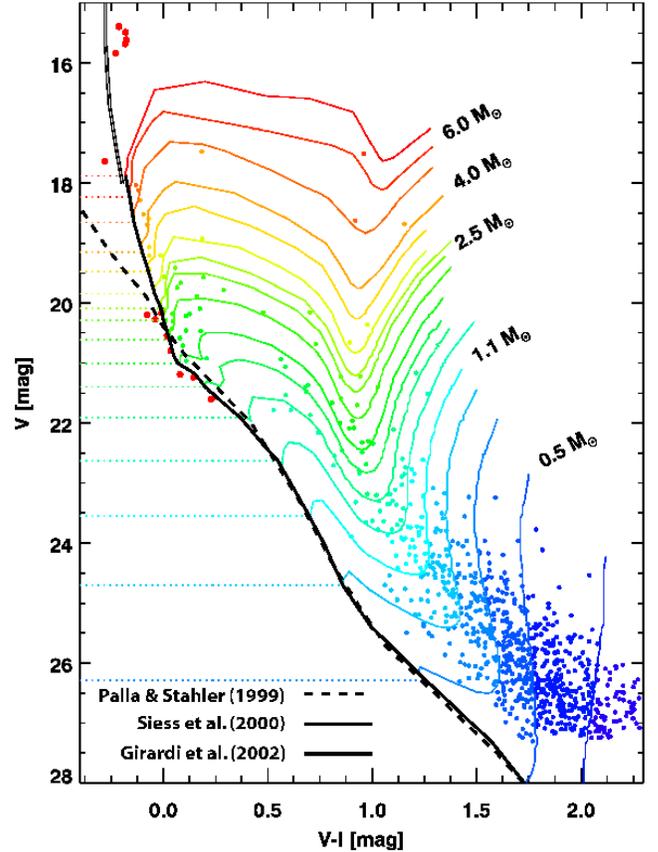}
\caption{Construction of the Initial Mass Spectrum (IMS) of NGC~602 by
counting stars between evolutionary tracks. The PMS tracks by Siess et
al. (2000) are shown in this plot. The ZAMS of both Siess et al. (2000)
and Palla \& Stahler (1999) are shown to demonstrate their differences
for stars with $V \lsim 20$~mag. The ZAMS from the grids of models by
Girardi et al. (2002), which is found to fit very well the one by Siess
et al. (2000) is plotted for the upper MS of the CMD. This ZAMS has been
used for the construction of a mass-luminosity relation for the bright
MS stars with $M \gsim 6$~M{\solar} for the derivation of their masses.
Different PMS tracks and the stars included between them are plotted in
different colors to be easier distinguished.\label{f:boxes}}
\end{figure}

As a consequence a ML relation cannot be used for the determination of
the masses of the observed PMS stars in NGC~602. Therefore, in the
following sections we study the IMS of the association, as we
constructed it by applying an age-independent approach of counting the
PMS stars within PMS evolutionary tracks (e.g. Gouliermis et al. 2006a).
For this method we use two different grids of evolutionary tracks, by
Palla \& Stahler (1999) and Siess et al. (2000), for stellar masses
between 0.1 and 6 M{\solar}. Both grids cover an age range starting at
the birth-line up to $\approx$~20~Myr, and they are available in the
standard Johnson-Cousins photometric system. In order to compare our CMD
to these models we transformed our ACS $m_{\rm 555}$ and $m_{\rm 814}$
magnitudes to the standard $V$ and $I$ respectively with the use of the
transformations provided by Sirianni et al. (2005; \S~8.3, and Appendix
D). In Fig.~\ref{f:boxes} we show the transformed $V-I$, $V$ CMD of
NGC~602 with the case of counting PMS stars between tracks from the
models by Siess et al. (2000). Although the lowest metallicity
considered for this grid of models is $Z=0.01$ ($\simeq$~2.5 times
higher than that of the SMC) it fits our data better than the one from
Palla \& Stahler (1999), which assumes the metallicity of the SMC of
$Z=0.004$. Both grids agree for low mass stars up to
$\approx$~2.5~M{\solar}. For higher masses the models by Palla \&
Stahler (1999) show a shift to the blue. This can be demonstrated by the
comparison of the ZAMS from both grids of models (Fig.~\ref{f:boxes}).
The most massive part of the IMS, for stars with $M$~\gsim~6~M{\solar},
is occupied by evolved MS stars, and therefore another grid of models
for such stars, was used. These models were compiled by Girardi et al.
(2002) for the standard photometric filter system and the metallicity of
the SMC. It should be noted that the ZAMS of Siess et al. (2000)
coincides better with the one from Girardi et al. (2002), which match
very well the observed MS of the system.

For the construction of the IMS we count all stars with a completeness
$\geq$~50\%, and therefore only stars with $m_{555}$~\lsim~24.5~mag,
corresponding to $M$~\gsim~1.1~M{\solar} are counted. Considering that
each evolutionary track corresponds to a specific mass, we constructed
the IMS by counting the PMS stars between evolutionary tracks based on
their loci in the CMD. We constructed thus, a histogram with bins
defined by the available PMS tracks. It should be noted that building a
histogram is the simplest, but by no means optimum, way of analyzing the
distribution of masses. Given the uncertainties in our photometry,
produced by the characteristics of the PMS stars, which translate into
corresponding mass uncertainties, it is possible that some of the
objects in every bin might actually have masses slightly outside the bin
range. Moreover, Ma{\'{\i}}z Apell{\'a}niz \& {\'U}beda (2005) found
significant numerical biases in the determination of the slope of power
laws from uniformly binned data using linear regression. These biases
are caused by the correlation between the number of stars per bin and
the assigned weights, and are especially important when the number of
stars per bin is small. This result implies the existence of systematic
errors in the values of IMFs calculated in this way. As an alternative,
those authors proposed to use variable-size bins and divide the stars
evenly among them. Such variable-size bins yield very small biases that
are only weakly dependent on the number of stars per bin. 

Taking into account the above, in order to avoid artificial features in
the distribution of the masses, we smoothed our counts by folding them
with an adaptive kernel function using a simple boxcar average procedure
of 3 mass-bins width. This process smoothes out any random fluctuations
due to the fact that some stars could belong to neighboring bins due to
the photometric uncertainties and inaccuracies of the evolutionary
tracks. Furthermore, in order to properly eliminate any numerical biases
in the estimation of the IMS slope due to the bin sizes, the stellar
numbers per mass-bin, $\tilde N$, which are used for the IMS, are the
counted incompleteness-corrected stellar numbers, $N$, but normalized to
a bin-size of 1~M{\solar}, by dividing them to the sizes $\Delta M$
of their corresponding bins.

\subsection{The Initial Mass Spectrum of NGC~602}
\label{s:ms}

\begin{figure}
\epsscale{1.15}
\plotone{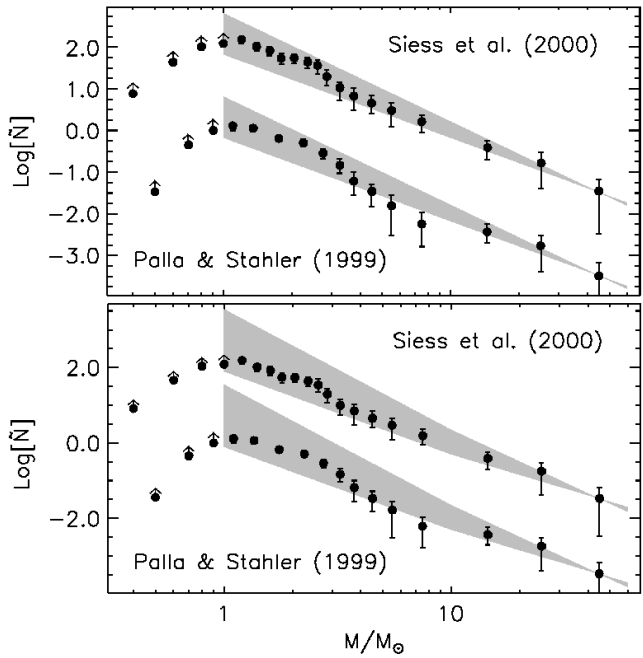}
\caption{The Initial Mass Spectrum (IMS) of NGC~602 as constructed by
counting PMS stars between evolutionary tracks and with the use of a ML
relation from the models by Girardi et al. (2002) for the brightest MS
stars. The derived IMS is shown in each panel from counting PMS stars
between tracks from both the models of Siess et al. (2000) (top IMS) and
Palla \& Stahler (1999) (bottom IMS). The shades at the top panel
correspond to a Kroupa (2002) IMS with $\alpha=-2.3\pm0.3$, while the
ones at the bottom panel to a Scalo (1998) one with $-2.7 \lsim \alpha
\lsim -2.3$. Stellar numbers $\tilde N$ are corrected to a bin-size of
1~M{\solar}. A slope of $\alpha\simeq-2.2\pm0.3$ is found to represent
very well the IMS as it is constructed with the use of all considered
grids of models.
\label{f:ims}}
\end{figure}

The constructed IMS of NGC~602 with the use of evolutionary tracks from
both Palla \& Stahler (1999) and Siess et al. (2000) is shown in
Fig.~\ref{f:ims}. Small changes in the slope of this IMS, which are
apparent in the figure, especially the ``knee''-like feature at
$\approx$~2.5~M{\solar}, imply that maybe a multi-power law is
appropriate to represent this IMS. However our modeling of the IMS with
a weighted linear fit proved that a single-power law of the form $\tilde
N\sim M^\alpha$ describes best this IMS. The slopes derived for the IMS
constructed with the use of models by Palla \& Stahler (1999) and Siess
et al. (2000) show that the use of both grids gives more or less the
same results within the derived uncertainties. Specifically, the derived
slope for the IMS constructed with the use of the tracks by Palla \&
Stahler (1999) is found to be $\alpha \simeq -2.0 \pm 0.3$, and the
slope derived with the use of the tracks by Siess et al. (2000) $\alpha
\simeq -2.4 \pm 0.2$ for the mass range 1~\lsim~$M/$M{\solar}~\lsim~45.
It should be noted that if we compare the derived IMS slopes for
specific narrower mass ranges, the largest difference between the used
models is seen in the lower mass range of 1~\lsim~$M$/M{\solar}~$<$~2.5
(although these differences are covered by the fitting errors). This is,
however, expected from the great differences between the ZAMS of Palla
\& Stahler (1999) and the one of Siess et al. (2000) shown in
Fig.~\ref{f:cmdngc}, which affects mostly stars with $M \gsim
2.5$~M{\solar}.
 
Taking into account the results from the application of the linear
regression for the IMS constructed with the use of both libraries of
evolutionary tracks, the IMS of NGC~602 is best described by the
function \begin{equation} \tilde N \sim M^{-2.2\pm0.3} \label{e:sie1}
\end{equation} for the whole observed mass range of $\sim$~1~M{\solar}
up to $\approx$~45~M{\solar}. The derived mean slope has a remarkable
coincidence with the slope of $\alpha=-2.35$ found by Salpeter (1955)
for the solar neighborhood, and with the results of Kroupa (2002) who
suggests a slope of $\alpha=-2.3\pm0.3$ for the mass range of
0.5~\lsim~$M$/M{\solar}~\lsim~100 as the average field star IMF. The IMS
of Kroupa (2002) is represented by the shaded region overplotted at the
top panel of Fig.~\ref{f:ims}.

A third IMS to be considered for comparing the IMS of NGC~602 is by
Scalo (1998) with a slope of $\alpha=-2.7\pm0.5$ for
1~\lsim~$M$/M{\solar}~\lsim~10 and $\alpha=-2.3\pm0.5$ for
10~\lsim~$M$/M{\solar}~\lsim~100. The slopes we find for the IMS of
NGC~602 are comparable also with this IMS for $M\gtrsim 10$~M{\solar}.
For smaller masses the slopes are more different, but still consistent
within the errors. The IMS of Scalo (1998) is represented by the shaded
region overplotted at the bottom panel of Fig.~\ref{f:ims}. It is worth
noting that our single-power law slope for the IMS of NGC~602 agrees
very well with the one found by Gouliermis et al. (2006a) for the stellar
association LH~52 in the Large Magellanic Cloud, $\alpha=-2.3\pm1.0$.
These authors could estimate the IMS slope only for a narrow mass range
(0.8~\lsim~$M$/M{\solar}~\lsim~1.4) due to observational limitations. A
multi-power law was found by Preibisch et al. (2002) for the IMS of the whole
Upper Sco OB association with the use of data from 2dF, {\em ROSAT}, and
{\em Hipparcos}. These authors found a slope of $\alpha=-2.8\pm0.5$ for
0.6~\lsim~$M$/M{\solar}~\lsim~2 and $\alpha=-2.6\pm0.5$ for
2~\lsim~$M$/M{\solar}~\lsim~20. Both these slopes are systematically
steeper than the one we found for NGC~602.

\subsection{Mass Spectra of Cluster~A and B~164}

\begin{figure}[t]
\epsscale{1.15}
\plotone{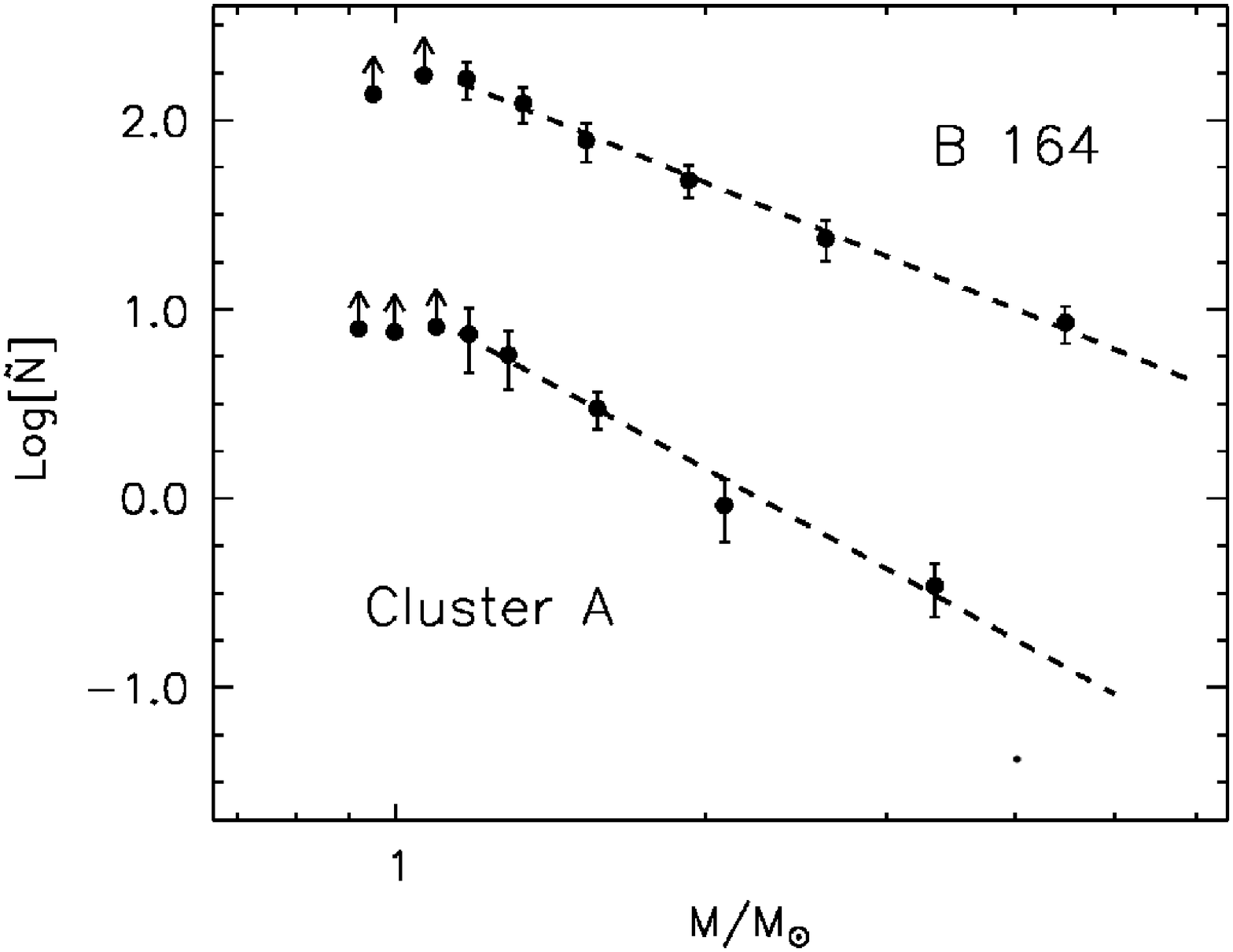}
\caption{The mass spectrum of both B~164 and Cluster~A, constructed
using ML-relations based on the evolutionary models by Girardi et al. (2002). 
Points derived from incomplete data (with completeness $<$~50\%) are
shown with arrows. The mass spectrum of both clusters is well
represented by a single-power law comparable to a Salpeter
IMS.\label{f:imfab}}
\end{figure}

As it is seen earlier from their CMDs, both B~164 and Cluster~A are open
clusters without any significant PMS population. Therefore, the
construction of their mass spectra is more straightforward, since it can
be done with the use of mass-luminosity (ML) relations. These relations
can be derived from certain models for evolved MS stars (Girardi et al.
2002), and they provide an accurate transformation of luminosities into
masses for the stars of these clusters. Naturally, the choice of the
appropriate isochrones, as the most representative for the age of the
systems, is crucial for this procedure. For this selection we used the
upper age limits found earlier (\S~\ref{s:pop}) to be $\approx$~160~Myr
for Cluster~A and $\approx$~80~Myr for B~164. After obtaining the masses
of the cluster members we constructed the corresponding IMS of each
cluster, based on the results of Ma{\'{\i}}z Apell{\'a}niz \& {\'U}beda
(2005), by counting the stars into variable-size mass-bins.

A single-power law is also found to represent the IMS of both clusters
for the whole observed mass range with $M \gsim$~1.5~M$_\odot$. The
best-fitting slopes of the corresponding IMS are found to be:
\begin{equation} \tilde N \sim M^{-2.2\pm 0.2}~~{\rm
for~B~164,}~~~~~\tilde N \sim M^{-2.9\pm 0.5}~~{\rm for~Cluster~A}
\label{e:pal} \end{equation} The IMS slope of Cluster~A is found
significantly steeper than that of B~164. Specifically, it is
interesting to note that the IMS of B~164 is more similar to the one by
Kroupa (2002) ($\alpha=-2.3\pm0.5$), whereas the IMS of Cluster~A fits
better to the one by Preibisch et al. (2002) (between $\alpha=-2.8$ and
$-2.6$ $\pm0.5$) or by Scalo (1998) ($\alpha=-2.7\pm0.5$). 


\section{CONCLUSIONS}
\label{s:conclusions}

We present our photometric study from data obtained in the filters
$F555W$ and $F814W$ with HST/ACS in wide-field mode of the star forming
region NGC~602/N~90 in the wing of the SMC. Our aim is to take advantage
of the high-resolution efficiency of ACS for the construction of the IMF
of the association NGC~602 for the whole observed stellar population of
the system. We use the ACS module of the photometric package DOLPHOT,
which is especially designed for imaging with the ACS and we provide the
full photometric catalog of all stars detected with short and long
exposures in one ACS/WFC field that covers a region of about $3\farcm4
\times 3\farcm4$ centered on the young stellar association NGC~602. The
region is found to host a mixture of stellar populations. Star counts
revealed that there are three prominent stellar concentrations in the
observed field: The association NGC~602 itself at the center, and two
young open clusters, B~164 (Bruck 1976), and one currently un-cataloged,
which we name ``Cluster~A'', both located at the northern edge of the
observed area. The CMDs of all three systems are contaminated by the
stellar population of the general field of SMC in the region. We
selected the most empty (and less contaminated from dust emission, as
seen with {\em Spitzer}; Gouliermis et al. 2007) part of the observed
area as the most representative of the general SMC field. We, then,
applied a Monte Carlo method for decontaminating the CMDs of all three
identified stellar systems from the contribution of the field.

The CMDs of B~164 and Cluster~A show features typical for young open
clusters, with a fully populated main sequence extending from the bright
part (with only a couple of bright stars in each cluster) down to the
detection limit of $m_{555} \sim 28$~mag. By comparing the brightest MS
stars of both B~164 and Cluster~A we were able to confine their maximum
age to be around 80~Myr and 160~Myr respectively. If we consider one red
super-giant in the CMD of Cluster~A as cluster member, then both
clusters have the same age and their close separation suggests that they
were possibly formed during the same star formation event.  On the other
hand, NGC~602 is a much younger system with some bright nebulosity still
in its vicinity. The CMD of the association reveals that the stellar
members of NGC~602 belong to a rich upper bright main
sequence\footnote{The 10 brightest stars in the system are identified as
early-type with spectral types between O5.5 and B0.5} down to $m_{555}
\sim 21$~mag and to a prominent red sequence of faint pre-main sequence
stars down to the detection limit. In general, the observed sequence of
PMS stars corresponds to an age of roughly 4~Myr. However, there is also
a significant amount of PMS stars that can be fitted by a $\sim$~10~Myr
model. The ``field'', the area selected as the most representative of
the general SMC field, is characterized by a well populated low main
sequence and a rather poor red giant branch with a turnoff at $m_{555}
\sim 22$~mag. Few stars brighter than the turnoff, and few PMS stars
were also included in this area, since it is located within the observed
FoV. 

We constructed the luminosity functions of all three stellar systems, by
counting the stars in magnitude bins of 0.5 mag in both $m_{555}$ and
$m_{814}$. Concerning the mass spectrum of the association, the PMS
stars in the CMD of NGC~602 show a prominent broadening, which does not
allow a direct measurement of the masses of these stars through a single
ML-relation. This broadening might be the result of an age spread, but
it could also be due to the characteristics of the PMS stars, such as
variability and extinction. In any case the accurate measurement of
their age and the estimation of their masses becomes a rather difficult
task.  Therefore, we constructed the initial mass spectrum of the
association with an age-independent method based on counting the PMS
stars between evolutionary tracks. We used the grids of models by both
Palla \& Stahler (1999) and Siess et al. (2000). For the bright main
sequence stars we used a ML-relation obtained from the evolutionary
models by Girardi et al. (2002). The IMS of the association is found to
be well represented by a single-power law, corresponding to an IMF of
slope $\Gamma \approx -1.2$ for 1~\lsim~$M$/M{\solar}~\lsim~45, similar
to the field IMF in the solar neighborhood (Salpeter 1955; Kroupa 2002).
No significant differences are found between the IMF derived with the
use of Palla \& Stahler (1999) models and the one from the models by
Siess et al. (2000), although these grids of models are found to be
quite different from each other for stars with $m_{555}$~\lsim~20~mag.
We also constructed the mass spectra of the clusters B~164 and
Cluster~A, and we found that the mass spectrum of B~164 corresponds to a
mass function similar to NGC~602 with $\Gamma \approx -1.2$, while the
one of Cluster~A is found to be steeper with $\Gamma \approx -1.9$.

The study of NGC~602/N~90 with observations from {\em Spitzer}/IRAC by
Gouliermis et al. (2007) showed evidence of ongoing star formation on
the rim of the {\sc H~ii} region. Candidate YSOs were identified by
these authors, located around the region of NGC~602, and their formation
is most likely triggered through a sequential event initiated at the
central association. This hypothesis is currently under investigation
within our study of the recent star formation history of this region
through the reconstruction of the observed sequence of PMS stars in and
around the association in synthetic CMDs (M. Schmalzl et al., in
preparation).

\acknowledgments

D. A. Gouliermis kindly acknowledges the support of the German Research
Foundation (Deu\-tsche For\-schungs\-ge\-mein\-schaft - DFG) through the
grant GO~1659/1-1. We would like to thank the unknown referee for
her/his constructive suggestions that helped to improve the manuscript
significantly. This paper is based on observations made with the
NASA/ESA {\em Hubble Space Telescope}, obtained from the data archive at
the Space Telescope Science Institute. STScI is operated by the
Association of Universities for Research in Astronomy, Inc. under NASA
contract NAS 5-26555. This research has made use of NASA's Astrophysics
Data System, Aladin (Bonnarel et al.\ 2000) and the SIMBAD database,
operated at CDS, Strasbourg, France.





Facilities: \facility{HST(ACS)}.




\begin{references}

\reference{} Alcaino, G., Alvarado, F., Borissova, J., \& Kurtev, R.\
2003, \aap, 400, 917

\reference{} Bonnarel, F., et al.\ 2000, \aaps, 143, 33

\reference{} Brandner W., Grebel, E. K., Barb\'{a}, R. H., et al. 2001,
AJ, 122, 858

\reference{} Brice{\~n}o, C., Calvet, N., Hern{\'a}ndez, J., Vivas,
A.~K., Hartmann, L., Downes, J.~J., \& Berlind, P.\ 2005, \aj, 129, 907

\reference{} Brice{\~n}o, C., Preibisch, T., Sherry, W.~H., Mamajek,
E.~A., Mathieu, R.~D., Walter, F.~M., \& Zinnecker, H.\ 2007, Protostars
and Planets V, 345

\reference{} Bruck, M.~T. 1976, Occasional Reports of the Royal
Observatory Edinburgh, 1, 1

\reference{} Davies, R.~D., Elliott, K.~H., Meaburn, J. 1976, \memras,
81, 89

\reference{} Dolphin, A.~E. 2000, PASP, 112, 1383

\reference{} Dolphin, A.~E., et al.\ 2001, \apj, 562, 303

\reference{} Fletcher, A.~B., \& Stahler, S.~W.\ 1994, \apj, 435, 329


\reference{} Girardi, L., et al.\ 2002, \aap, 391, 195

\reference{} Gouliermis, D., Brandner, W., \& Henning, T.\ 2005, \apj,
623, 846

\reference{} Gouliermis, D., Brandner, W., \& Henning, T.\ 2006a, \apjl,
636, L133

\reference{} Gouliermis, D.~A., Dolphin, A.~E., Brandner, W., \&
Henning, T.\ 2006b, \apjs, 166, 549

\reference{} Gouliermis, D.~A., Quanz, S.~P., Henning, T. 2007, ApJ,
665, 306

\reference{} Henize, K.~G. 1956, ApJS, 2, 315

\reference{} Hennekemper, E., Gouliermis, D.~A., Henning, T., Brandner,
W., Dolphin, A. E. 2008, ApJ, 672, 914

\reference{} Herbst, W., Herbst, D. K., Grossman, E. J., \& Weinstein, D.
1994, AJ, 108, 1906

\reference{} Herbst, W., et al. 2002, A\&A, 396, 513

\reference{} Hester, J.~J., et al.\ 1996, \aj, 111, 2349

\reference{} Hutchings, J.~B., Cartledge, S., Pazder, J., \& Thompson,
I.~B.\ 1991, \aj, 101, 933

\reference{} Hunter, D.~A., Elmegreen, B.~G., Dupuy, T.~J., \&
Mortonson, M.\ 2003, \aj, 126, 1836

\reference{} Koekemoer, A.~M., Fruchter, A.~S., Hook, R.~N., Hack, W.
2002, The 2002 HST Calibration Workshop : Hubble after the Installation
of the ACS and the NICMOS Cooling System, Proceedings of a Workshop held
at the Space Telescope Science Institute, Baltimore, Maryland, October
17 and 18, 2002.~ Edited by Santiago Arribas, Anton Koekemoer, and Brad
Whitmore.~Baltimore, MD: Space Telescope Science Institute, p.337

\reference{} Koornneef, J. 1983, \aap, 128, 84

\reference{} Kontizas, E., Kontizas, M., Gouliermis, D., Dapergolas, A.,
Korakitis, R., Morgan, D.~H. 1999, IAU Symposium 190, New Views of the
Magellanic Clouds, eds. Y.-H. Chu et al., 410

\reference{} Kroupa, P.\ 2002, Science, 295, 82

\reference{} Luck, R.~E., Moffett, T.~J., Barnes, III, T.~G., Gieren,
W.~P. 1998, AJ, 115, 605

\reference{} Ma{\'{\i}}z Apell{\'a}niz, J., \& {\'U}beda, L.\ 2005,
\apj, 629, 873

\reference{} Massey, P.\ 2006, The Local Group as an Astrophysical
Laboratory, 164

\reference{} McCumber, M.~P., Garnett, D.~R., Dufour, R.~J. 2005, AJ,
130, 1083

\reference{} Miller, G.~E., \& Scalo, J.~M. 1979, \apjs, 41, 513

\reference{} No{\"e}l, N.~E.~D., Gallart, C., Costa, E., \& M{\'e}ndez,
R.~A.\ 2007, \aj, 133, 2037

\reference{} Nota, A., et al.\ 2006, \apjl, 640, L29

\reference{} Palla, F.\ 2005, Massive Star Birth: A Crossroads of
Astrophysics, 227, 196

\reference{} Palla, F., \& Stahler, S.~W.\ 1999, ApJ, 525, 772

\reference{} Panagia, N., Romaniello, M., Scuderi, S., \& Kirshner,
R.~P.\ 2000, ApJ, 539, 197

\reference{} Piatti, A.~E., Sarajedini, A., Geisler, D., Gallart, C., \&
Wischnjewsky, M.\ 2007, \mnras, 382, 1203

\reference{} Preibisch, T., Brown, A.~G.~A., Bridges, T., Guenther, E.,
\& Zinnecker, H.\ 2002, \aj, 124, 404

\reference{} Rochau, B., Gouliermis, D.~A., Brandner, W., Dolphin,
A.~E., \& Henning, T.\ 2007, \apj, 664, 322

\reference{} Romaniello, M., Scuderi, S., Panagia, N., et al. 2006,
A\&A, 446, 955

\reference{} Sabbi, E., et al.\ 2007, \aj, 133, 44 (Erratum: 2007, \aj,
133, 2430)

\reference{} Salpeter, E.~E.\ 1955, \apj, 121, 161

\reference{} Scalo, J.\ 1998, The Stellar 
Initial Mass Function (38th Herstmonceux Conference), 142, 201 

\reference{} Sherry, W.~H., Walter, F.~M., \& Wolk, S.~J.\ 2004, \aj,
128, 2316

\reference{} Siess, L., Dufour, E., Forestini, M. 2000, \aap, 358, 593

\reference{} Sirianni, M., Nota, A., Leitherer, C., De Marchi, G., \&
Clampin, M.\ 2000, \apj, 533, 203

\reference{} Sirianni, M., Nota, A., De Marchi, G., Leitherer, C., \&
Clampin, M.\ 2002, ApJ, 579, 275

\reference{} Sirianni, M., et al.\ 2005, \pasp, 117, 1049

\reference{} Stahler, S.~W., \& Fletcher, A.~B.\ 1991, Memorie della
Societa Astronomica Italiana, 62, 767

\reference{} Stanimirovic, S., Staveley-Smith, L., van der Hulst, J.~M.,
Bontekoe, T.~R., Kester, D.~J.~M., Jones, P.~A. 2000, MNRAS, 315, 791

\reference{} Venn, K.~A. 1999, ApJ, 518, 405

\reference{} Westerlund, B.~E.\ 1964, MNRAS, 127, 429

\reference{} Zinnecker, H.\ 1998, Highlights in Astronomy, 11, 136

\end{references}
\end{document}